\documentstyle[preprint,aps]{revtex}                                          
\begin{document}
\draft
\title{Relaxation oscillations in model sandpiles}
\author{J.E.S. Socolar and M.E. Bleich}
\address{Department of Physics and Center for Nonlinear and Complex Systems,
	 Duke University, Durham, NC 27708}
\date{\today}
\maketitle
\begin{abstract}
We introduce a simple one-dimensional sandpile model 
that undergoes relaxation oscillations.
A single model can account for self-organized critical behavior
and relaxation oscillations, depending on the manner in which
it is driven, mirroring the experimental situation for real sandpiles.
The relaxation oscillations are robust with respect to minor
modifications of the avalanche rules, including the application
of probabilistic rules.
\end{abstract}


Many extended dissipative
systems exhibit what may be called ``avalanche dynamics'',
meaning that they respond
to slow external driving by undergoing 
rapid, discrete relaxation events, or avalanches.
In general, such systems involve a quantity $\phi$
that is increased slowly by some external mechanism.
When $\phi(x)$ or $\nabla\phi(x)$ becomes too large, 
an avalanche is initiated as $\phi$ is redistributed around
a locally unstable region, possibly causing a chain reaction
of local instabilities.
The system then rapidly relaxes into a (temporarily) stable
state in which $\Phi$, the volume integral of $\phi(x)$,
has been reduced either by transport across the boundary
of the system or by nonconservative dynamics in the bulk.
This paper is concerned with the type of macroscopic
behavior that might be observed in such systems,
particularly the question of whether global relaxation
oscillations should be expected.

One system that has received much attention recently
is the proverbial sandpile.
Here $\phi(x)$ is the height of the pile 
at position $x$ and the instability
occurs when the local slope of the pile becomes too large.
``Avalanches'' consist of rearrangements on
the surface of the pile corresponding to the ordinary
meaning of the term, some of which transport mass
off of an open boundary of the pile.
Another example is the relaxation of the earth's crust
via earthquakes.
Here $\phi(x)$ is the stress, which can be
relieved by transport across an open boundary or by
nonlinear processes that do not conserve $\Phi$.

Perhaps the most obvious feature of
systems governed by avalanche dynamics that requires explanation
is the distribution of avalanche sizes, $P(s)$, where $s$ is
the total amount of $\Phi$ 
(mass, in the case of sandpiles) that leaves the system.
By analogy with equilibrium statistical mechanics,
one might expect that the qualitative features of $P(s)$ do
not depend on microscopic details of the avalanche dynamics,
but only on general features of the system such as its
symmetries.
If so, it should be possible to construct highly
simplified numerical models that show behavior similar
to real experiments on sand, just as the behavior of
the Ising model accurately reflects the general features
of a variety of microscopically complicated physical systems.

Following the seminal work of Bak et al. on self-organized
criticality in a class of numerical models \cite{baka}, 
Kadanoff et al.
introduced several 1D models that could be taken as
candidates for the essential dynamics of sandpiles. \cite{kada}
The models introduced are extremely simple
to write down, but show highly nontrivial behavior.
By analogy with equilibrium systems, one would expect
the generic behavior of $P(s)$ in the 1D system
to be one of two possibilities,
perhaps depending on the value of a system parameter:
(1) $P(s)$ decays exponentially beyond some size $s$
associated with a correlation length in the steady state or,
(2) $P(s)$ consists primarily of a large peak on the
order of $L^2$, where $L$ is the length of the system,
in which case the macroscopic slope of the pile would undergo 
oscillations whose amplitude does not decrease to zero 
in the infinite system-size limit; 
the macroscopice slope exhibits ``relaxation oscillations''.
The former possibility indicates a characteristic
size for avalanches, independent of $L$, while the latter
corresponds to a state dominated by huge avalanches that sweep
away a finite fraction of the mass of the pile.

The results of simulations of a variety of sandpile models
were remarkable in that
although stationary slope profiles
were attained and there were no relaxation oscillations,
$P(s)$ was found to be a broad distribution in which
no characteristic length scale other than some fractional power
of the system size could be identified.
In this sense, the models exhibit self-organized criticality (SOC),
tuning themselves to a ``critical point''
between possibilities (1) and (2) by virtue of their 
own internal dynamics.

Stimulated in part by the apparent ubiquity of SOC in toy
models, experiments have been performed on sandpiles
in various configurations.
The results to date remain inconclusive:
while there is some evidence of SOC in heaps formed
by grains dropped one at a time onto a flat plate \cite{held},
experiments employing a rotating drum half-filled with
sand find relaxation oscillations \cite{nagel}.

Taking the empirical evidence for relaxation oscillations
at face value, it appears that the numerical models previously
studied all lacked some crucial ingredient that is present
in real sandpiles.
As a first step in discovering what that ingredient is, we 
consider here a class of models obtained by introducing
modifications to the limited local sandpile (LLS) of Kadanoff et al.
We find that relaxation oscillations do occur under appropriate
conditions, as detailed below.
Our conclusion is that the models originally studied are
artificial in that they do not incorporate generic effects
that turn out to be relevant for the dynamics.
Studies of our more realistic model should yield
insights more generally applicable to real physical situations. 

The LLS consists of a set
of integer heights, $h_i$, defined on a 1D lattice $1\leq i\leq L$.
The avalanche dynamics are extremely simple: whenever
the slope $z_i \equiv h_i - h_{i+1}$ exceeds 2,
two units of height, called ``grains'', 
are transferred from $i$ to $i+1$.
Site $L+1$ is an open boundary; grains transferred to it
disappear and $h_{L+1}$ is always $0$.
The LLS is driven by the slow addition of individual grains
at randomly chosen sites, where ``slow'' means that
after the addition of a grain, any avalanche that occurs
is run to completion before the next grain is added.
When we refer to a ``state'' below, we mean the stable
configuration reached after a driving event.
A variation of the LLS, called the ``unlimited local sandpile'' \cite{kada}
(ULS) is the same except that when $z_i > 2$ the number of
grains transferred is $z_i -1$.

Our first observation is that the driving mechanisms used in
the experiments differ dramatically, so the effects
of different types of driving on the models must be
explored.
In particular, though the heap experiments do involve
the addition of grains more or less as specified in the model,
the rotating cylinder experiments involve uniform increases
in the slope of the surface of the sand.
A better representation of the latter is to drive the
LLS or ULS by adding a unit of {\em slope} to a
randomly chosen site rather than a unit of height.
We will call models driven in these different ways 
``slope driven'' and ``height driven''.

The height driven LLS and ULS are known to evolve
to self-organized critical states. \cite{kada}
Though both are quite subtle in detail, exhibiting
multifractal scaling properties,
$P(s)$ for finite size
systems clearly decays exponentially
for $s > cL$, where $c$ is a constant of order unity;
the largest avalanche sizes observed in
both models are of order $L$.   

The slope driven LLS turns out to be trivial.
One can easily see that it will eventually reach a state
in which $z_i > 0$ for all $i$ and that all subsequent states
will have the same property.
If slope is then added to a site with $z_i = 1$ there is
no avalanche.
If slope is added where $z_i = 2$, the state reached will
be identical to the original except that $z_i \rightarrow 1$
and the number of grains falling of the pile during the avalanche
is just $2i$.
Thus $P(s)$ is independent of $s$ for even $s < 2L$.
As in the height driven LLS, avalanches bigger than $2L$ 
can never occur and relaxation oscillations are impossible.

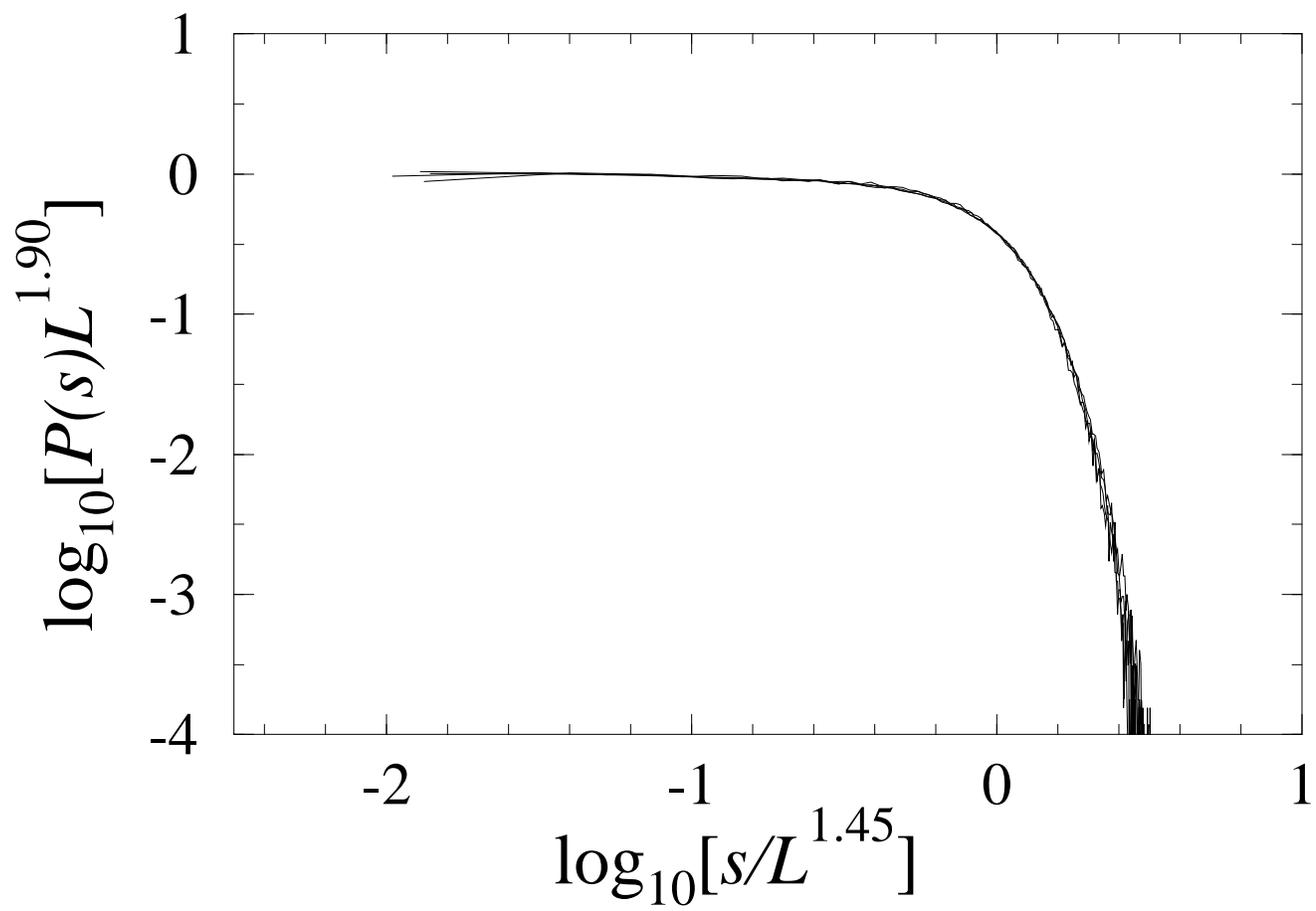
\begin{figure}
 \caption{Avalanche size distributions for the slope driven ULS.
	  Data are shown for $L=250$, 500, 1000, and 2000.
	  Each curve represents $\sim 5\times 10^{5}$ avalanches and was
	  obtained from $L\times 10^{4}$ driving events.
	  Each point plotted represents an average over a fixed binwidth
	  selected so that the curve contains $\sim 150$ points.}
 \label{fULS}
\end{figure}

The slope driven ULS is more interesting.
As in the LLS, once a state with all $z_i > 0$ is
reached, the system will always remain in such a state.
For a state $z_1,z_2,z_3,...z_i=2,...z_L$, 
making an addition of slope at $i$  
results in the state
$z_{i+1},...z_L,1,1,...1$.
The size of the avalanche in any particular case is easily calculated 
and for the case $z_j=2$ for all $j$ one finds $s = i(2L-i+1)/2$.
Thus it is possible to have an avalanche
size of order $L^2$, unlike in the LLS.
Nevertheless, numerical evidence indicates that $P(s)$
decays exponentially for $s > cL^{1.45}$.
(See Figure~\ref{fULS}. \cite{betanu})
This shows that the scaling properties of the system can
be strongly affected by the form of the driving, but also
that the slope driven ULS does not undergo relaxation oscillations.

We now introduce a new slope driven model for which $P(s)$ is dominated
by a peak at $s\sim L^2$.
The model is a variation of the LLS, which will be called
the ``dynamic, limited, local sandpile'' (DLLS),
designed to allow large avalanches to form by changing
the local stability criteria for a site that has already
participated in a given avalanche.
In order to state the rules of the DLLS, it is useful
to rephrase the LLS rules as follows:
When a grain is added at site $i$ making $z_i = 3$,
an avalanche is started.
Let $j<i$ be the closest trap to the left of $i$
and $k>i$ be the closest trap to the right of $i$,
where a trap is defined as a site with $z\leq 0$.
The net result of the avalanche is that two grains per
site are transferred from the sites $i$ through $j+1$
to the sites $k-(i-j)+1$ through $k$.
We say that a cluster of grains is destabilized,
slides down the pile until its front reaches the first
trap, and stops there.

The modification introduced in the DLLS is
that sites with $z=2$ are destabilized when
the back of a cluster slides over them,
and the 0 produced when two grains are
removed from such a site is not counted as a trap
for the duration of the avalanche.
More precisely: an initial destabilized cluster is
formed exactly as in the LLS;
it then advances down the pile (to the right) one site at a time;
if the last two grains in the cluster are moved
from a site $a$ with $z_a=2$, two grains are taken from $a$
and they, along with two grains per site between
$a$ and the closest trap to the left, are added to the back of
the sliding cluster;
though $z_a$ is now 0, it does not act as a trap
until the current avalanche is completed;
the avalanche stops when the front reaches a trap.
A site with $z=0$ that is not acting as a trap
will be called a ``proto-trap'' since it will become
a trap upon the completion of the avalanche.

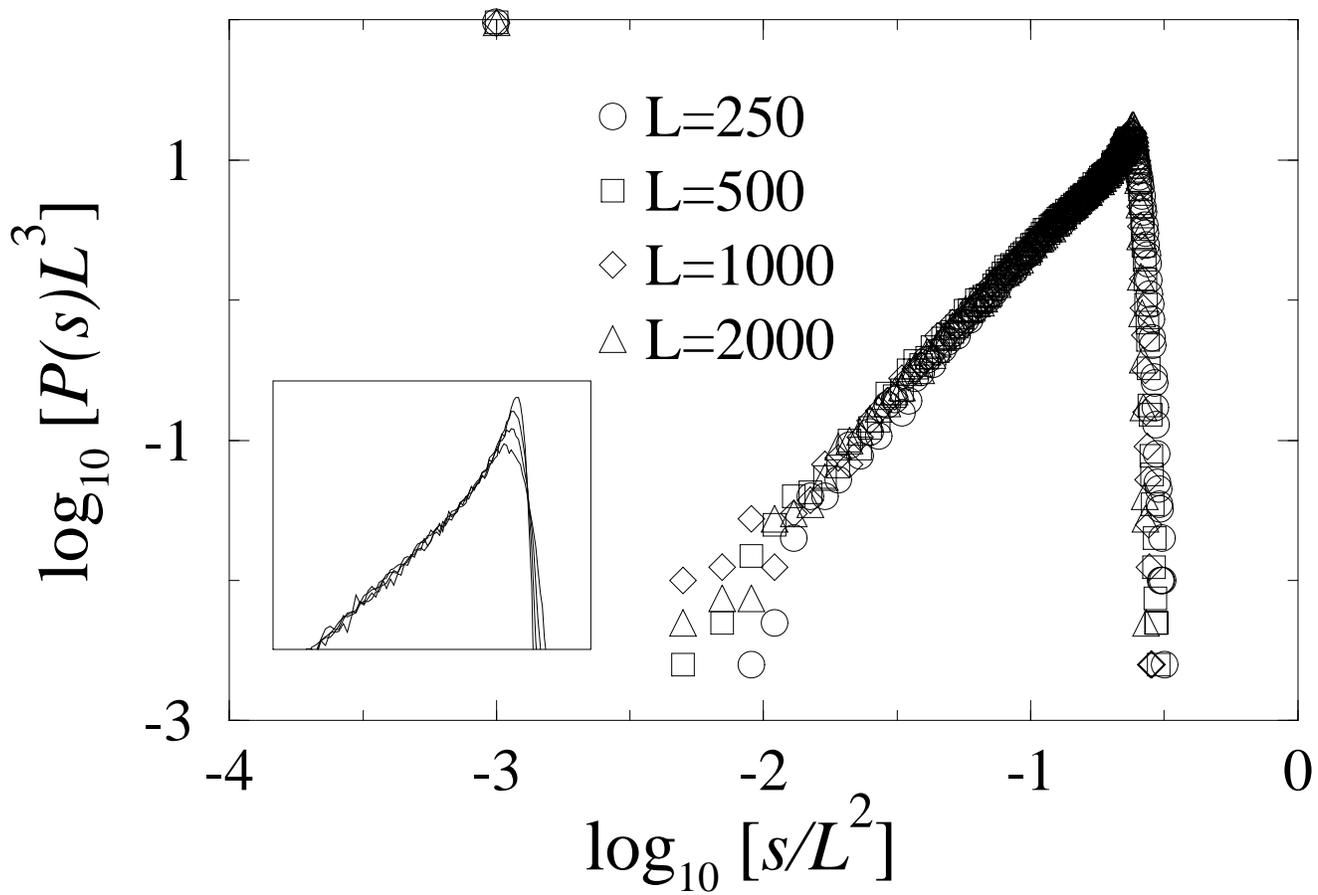
\begin{figure}
 \caption{Avalanche size distributions for the slope driven DLLS.
	  Data are shown for $L=250$, 500, 1000, and 2000.
	  Each curve represents $\sim 3\times 10^{5}$ avalanches and was
	  obtained from $2L\times 10^{5}$ driving events.
	  Each point plotted represents an average over a fixed binwidth
	  selected so that the curve contains $\sim 150$ points.}
 \label{fDLLSsd}
\end{figure}

The avalanche size distribution for the 
{\em slope driven} DLLS is shown in Figure~\ref{fDLLSsd}
for various system sizes.
\cite{sdefn}
For large $s$,
the data are well-fit by the scaling form 
$P(s) \sim L^{-\beta}\,f(s/L^\nu)$ with 
$\nu=2$ and $\beta=3$.\cite{betanu}
It is clear that the position of the peak at
$s/L^{2}\simeq 0.25$ does not shift, though its
shape changes near the tip and the trailing
edge is sharper for larger $L$ (see inset). 
This implies that individual avalanches remove
a finite fraction of the mass of the pile and 
that the average slope of the slope driven DLLS undergoes
relaxation oscillations.

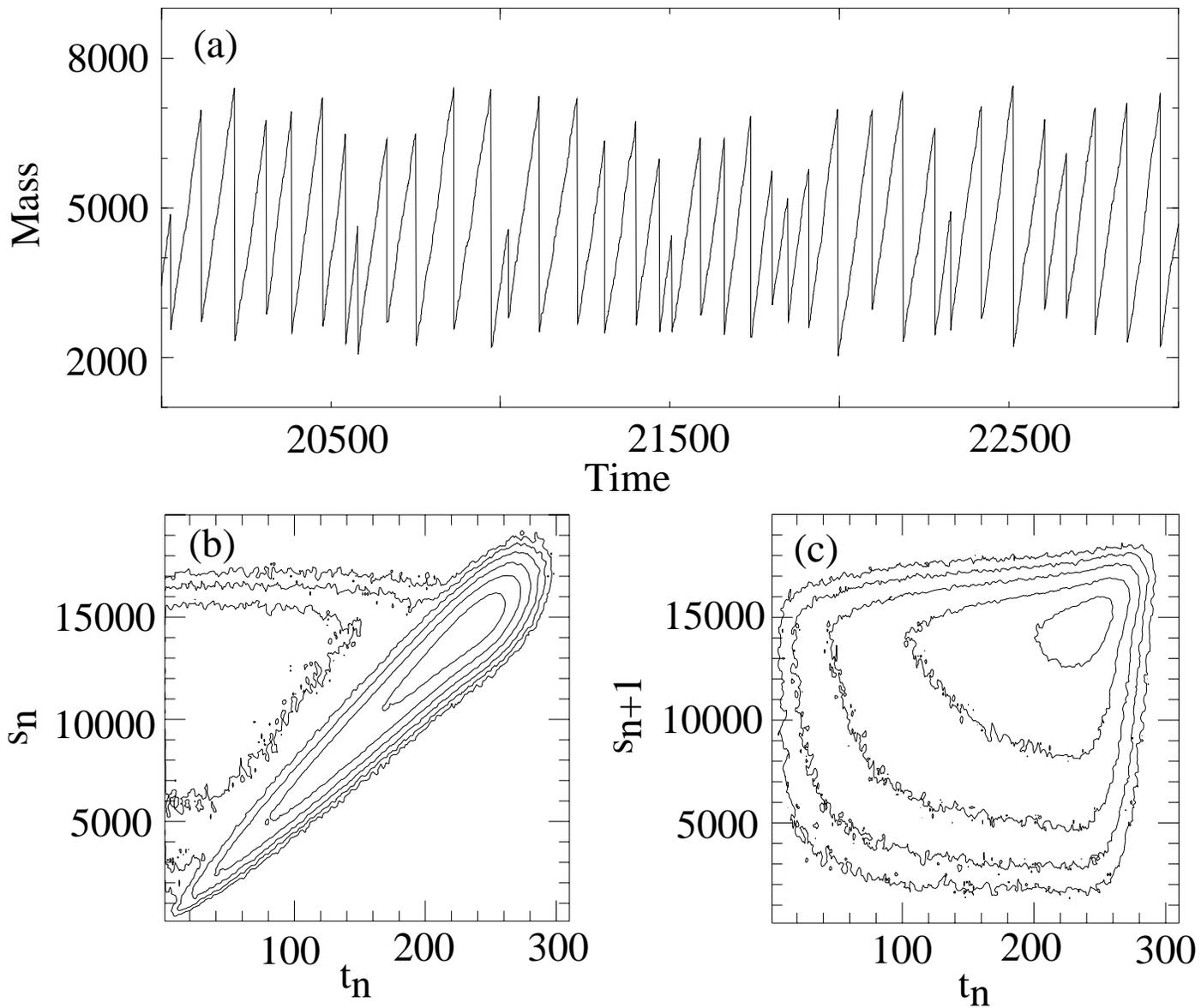
\begin{figure}
 \caption{(a) Total mass of the pile for the slope driven DLLS.
	  (b) $P(s_{n}, t_{n})$ (see text).
	  (c) $P(s_{n}, t_{n+1})$.
	  Each contour line represents a change in $P$ by a factor of 3.
	  Values of $s$ are binned as in Figure~2 and data was smoothed
	  by averaging over $3\times 3$ neighborhood.}
 \label{fmtDLLSsd}
\end{figure}

A plot of the total mass of the pile as a
function of the number of drops is shown 
in Figure~\ref{fmtDLLSsd}a.
A feature of interest is the correlation between the
size, $s_n$, of the $n^{th}$ avalanche and the time, $t_n$, 
(measured in numbers of driving events, or ``drops'') 
between it and the preceeding avalanche.
Figure~\ref{fmtDLLSsd}(b) shows a contour plot of 
the probability distribution $P(s_n, t_n)$,
which clearly indicates a correlation. \cite{contours}
In contrast, a plot of $P(s_n, t_{n+1})$ shows
a much weaker correlation;
the width of the distribution for fixed $s$ or $t$
is significantly larger in (c) than in (b).
This indicates that the large avalanches reset
the average slope of the pile to some fixed
reference value, but are triggered at a fairly
broad distribution of average slopes.
In the language used to describe fault dynamics,
the avalanches in this model are ``size-predictable''
rather than ``time-predictable'' \cite{scholz};
i.e., the time elapsed since the last avalanche
is useful information in predicting the size of
the next one, but knowing the size of the last
one does not help in predicting the time of
the next one.

We note that the relaxation oscillations are robust
with respect to the introduction of probabilistic
rules.
We have considered a modification in which
each time a site with $z=2$ would be destabilized
in the DLLS, the destabilization occurs only with
probability 1/2 and, furthermore, the probability
that the resulting $z=0$ site will be a proto-trap
rather than a real trap is 1/2.
The modified model also shows clear relaxation
oscillations for system sizes up to 1000.

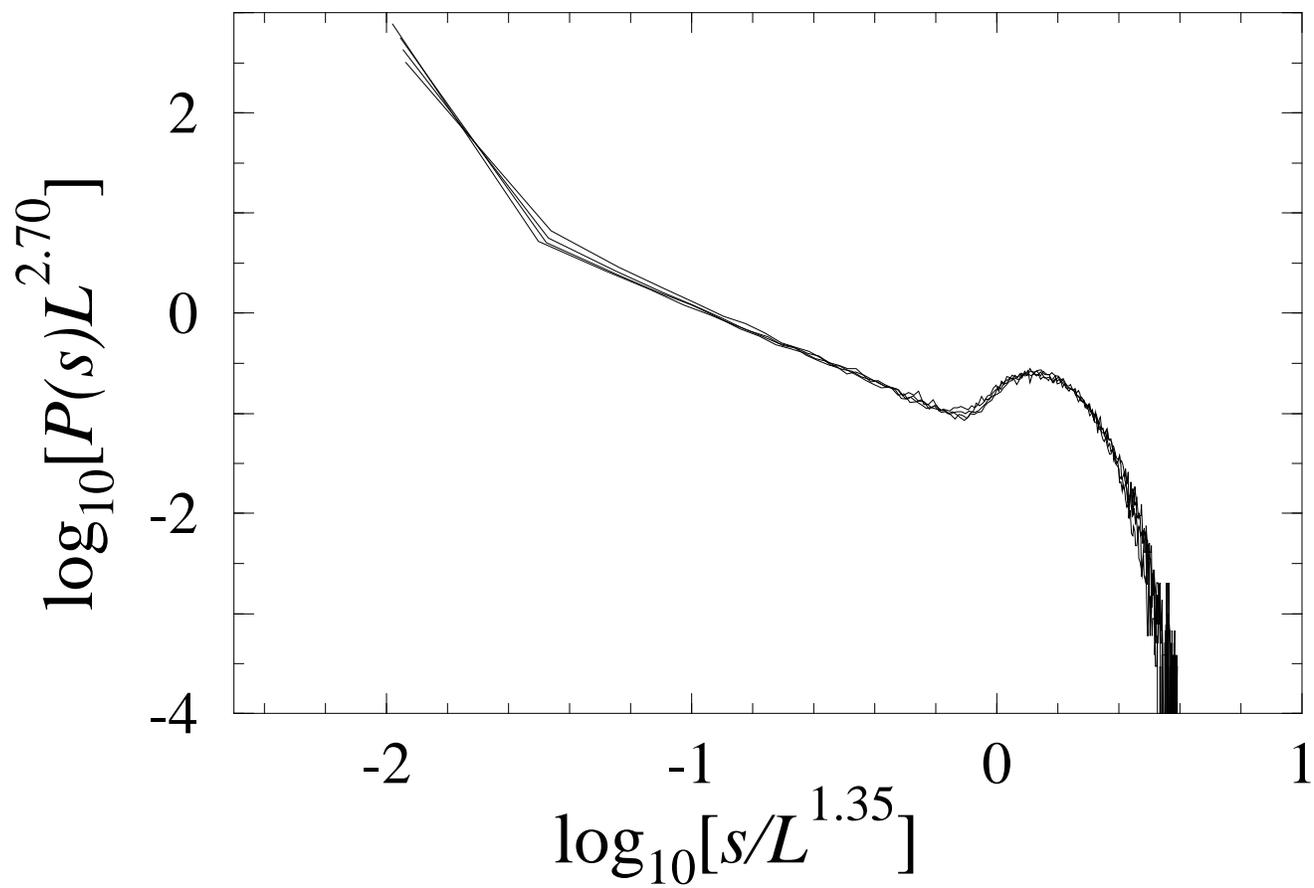
\begin{figure}
 \caption{Avalanche size distributions for the height driven DLLS.
	  Data are shown for $L=250$, 500, 1000, and 2000.
	  Each curve represents $\sim 1.2\times 10^{6}$ avalanches and was
	  obtained from $L\times 10^{6}$ driving events.
	  Each point plotted represents an average over a fixed binwidth
	  selected so that the curve contains $\sim 150$ points.}
 \label{fDLLShd}
\end{figure}

The {\em height driven} DLLS behaves quite differently,
exhibiting self-organized criticality rather than
relaxation oscillations.
Figure~\ref{fDLLShd} shows $P(s)$ for the height driven model
for various system sizes. \cite{betanu}
A reasonable data collapse is achieved by the same
scaling form used above but with $\nu=1.35$ and $\beta=1.70$.
In this case the average slope of the pile converges
to a stationary value in the infinite system limit.
Examination of the stationary slope profile reveals,
however, that the mechanism of selection of the critical
state is {\em not} the singular diffusion mechanism
discussed by Carlson et al. \cite{cara}. 
The profile does not exhibit a power-law approach
to a critical value at the open end of the system.
(Preliminary results show
a power-law convergence from {\em above} to some reference slope
as the {\em closed} boundary is approached.
Details will be discussed elsewhere.)
 
While it is too early to draw firm conclusions about
the relevance of the DLLS to real sandpiles,
there is a highly suggestive correspondence
between the qualitative behavior observed in the
model and the experiments of Held et al. \cite{held}
and Jaeger et al. \cite{nagel}
In the former, where individual grains were added at
random positions spanning the suface of a heap,
critical scaling was observed, as in the
height driven DLLS.
In the latter, a half-filled cylinder was slowly
rotated so that the slope of the sand surface slowly
increased everywhere.
In the slope driven DLLS, this process is modeled 
by a discrete process in which the sequence in
which the local slopes cross threshold values
is determined randomly, but the basic fact that
no local slope is decreased by the driving mechanism
is faithfully represented.
Both the experiment and the model show relaxation oscillations.

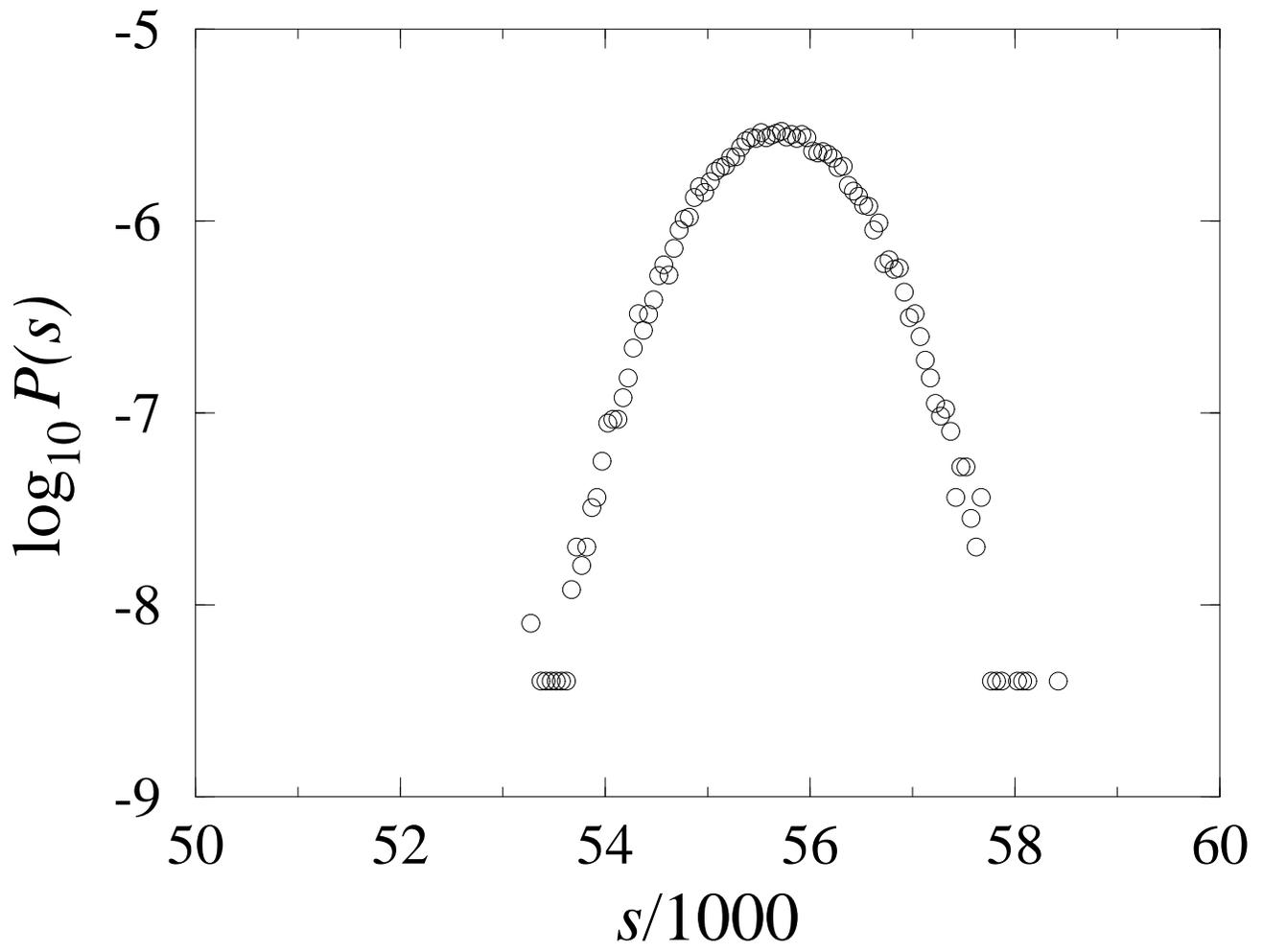
\begin{figure}
 \caption{Avalanche size distributions for the height driven DLLS
	  with drops performed on top half of pile only.
	  $L=500$.  Note the expanded horizontal scale, which includes
	  all nonzero data points except $s=0$.
	  Each point plotted represents an average over a fixed binwidth
	  selected so that the curve contains $\sim 150$ points.}
 \label{fDLLShdhalf}
\end{figure}

Furthermore, in experiments done by Held et al. on
larger heaps in which the random positions where grains were
added spanned only the top half of the pile, a power law
distribution was not observed, but rather a distribution
dominated by large avalanches. \cite{heldu}
Remarkably, this also occurs in the height driven DLLS, as evidenced
by the distribution shown in Figure~\ref{fDLLShdhalf}.
(The manner in which this distribution scales with increasing
$L$ depends on what is held constant, the fraction of the
pile covered by drops, the distance from the top that is covered,
or the distance from the bottom that is uncovered.)

In an effort to determine which features of the DLLS are essential
for generating relaxation oscillations, we have investigated
one further model, the ``dynamic, unlimited, local sandpile'' (DULS),
which is a modification of the ULS in which sites that
topple have a reduced toppling threshold for two lattice updates
during an avalanche.  
(Details of the rules will be published elsewhere.)
We have found that this dynamic effect does {\em not}
produce relaxation oscillations in the slope driven model,
though it does increase the exponent $\nu$ significantly.
(Preliminary results give $\nu\simeq 1.75$.)

We conjecture that the ingredients crucial for
generating relaxation oscillations are: 
(1) rules that allow large avalanches, which required
the dynamic modification of the LLS, but would not
require modification of the ULS;
(2) slope driving; and
(3) rules that generate effective traps in the wake 
of a large avalanche,
an obvious feature of the DLLS that is not
present in the ULS or DULS.

The coincidence between the qualitative behaviors observed in
the DLLS and real sandpiles suggests that models of this type
may yield useful insights into avalanche dynamics in generic
physical systems.
The question remains, however, as to whether the effects included
in the DLLS are somehow artificial.
In our view, the primary issue at this point is the
temporal nonlocality implicit in the rules for proto-traps.
As currently defined, the model requires that a
proto-trap not turn into a real trap until the entire
avalanche is completed.
This may not be unreasonable, given the required separation
in time scales between the avalanche dynamics and the driving rate.
Nevertheless, it is important to find out whether
a system governed by strictly local rules
can produce the same behavior.
In any case, investigation of the DLLS is significant
in that it highlights certain features that can 
affect the qualitative macroscopic behavior of
a broad class of slowly driven, dissipative systems.

\begin{references}
\bibitem{betanu} We fit $P(s)$ to the finite-size scaling form 
		$L^{-\beta}f(s/L^{\nu})$. 
		$P(s)$ is the probability of an 
		avalanche of size $s$ occuring per driving event.
		With this normalization the sum rule given in
		Eq.~(2.15) of Ref.~\cite{kada} reads
		$\int sP(s)ds = 1$ (height driving) or $L/2$ 
		(slope driving), implying 
		$\beta=2\nu$ or $2\nu - 1$, respectively.
\bibitem{sdefn} For the DLLS, the avalanche size $s$ is defined as
		half the number of grains falling off the pile because
		the grains always fall off in pairs.
\bibitem{contours} In Figure~\ref{fmtDLLSsd}(b), 
		the nonzero density to the left of the
		main peak is due to the avalanches in the smallest
		bin in Figure~\ref{fDLLSsd}.
		An analogous plot in which only avalanche sizes
		greater than $L^{2}/500$ and the elapsed times
		between such events are considered consists of
		a clean single peak along the diagonal.
\end{references}
\end{document}